# Notes on relaxation of the resonant model using a scattering approach

*Alexander Semenov*

*231 S 34 St., Department of Chemistry, University of Pennsylvania, Philadelphia, Pennsylvania, 19104*

Email: *asemenov@sas.upenn.edu*

Abstract:

Here we consider a non-perturbative description of the non-interacting resonant model using a scattering approach employed in JCP 152(24) 244126 (2020). We showed that such description coincides with the standard NEGF and Landauer-Buttiker approaches

## 1. Introduction.

Suppose that a system of infinite size (say, baths connected through a central region) is characterized by a steady-state density matrix $\hat{\rho}_0$ and the total Hamiltonian of the system is $\hat{H}_0$. Suppose that the Hamiltonian has undergone a sudden perturbation $\hat{V}$ at some time $T_1$. What will happen with the density matrix?

The boundary condition:

$$\hat{\rho}(t = T_1) = \hat{\rho}_0 . \tag{1.1}$$

Thus

$$\hat{\rho}(t = T_2) = \exp(-i\hat{H}(T_2 - T_1)/\hbar)\hat{\rho}_0 \exp(i\hat{H}(T_2 - T_1)/\hbar) \tag{1.2}$$

where

$$\hat{H} = \hat{H}_0 + \hat{V} \tag{1.3}$$

In Ref. 1 we introduced the following operator

$$\hat{\tilde{\Omega}}(T_2, T_1) = \exp\left(\frac{-i}{\hbar}\hat{H}(T_2 - T_1)\right) \tag{1.4}$$

We proved that the following expression

$$\hat{\rho} = \hat{\tilde{\Omega}}(0, -\infty)\hat{\rho}_0 \hat{\tilde{\Omega}}^\dagger(0, -\infty) \tag{1.5}$$

corresponds to a steady state density matrix for the Hamiltonian $\hat{H} = \hat{H}_0 + \hat{V}$ i.e. $[\hat{\rho}, \hat{H}] = 0$. Since $\hat{\tilde{\Omega}}(0,-\infty) = \hat{\tilde{\Omega}}(\infty,0) = \hat{\tilde{\Omega}}(\infty,T_1)$ at, $t \to \infty$ the matrix $\hat{\rho}_0$ will relax to a new steady state $\hat{\rho}(t = \infty)$.

This conclusion seems a bit off: how could a unitary transformation give a relaxation? The answer is that a unitary transformation for *a finite size* system does not produce any relaxation. That is not true for infinite size system. To prove this more "explicitly", consider a sudden "kick" of the dot level in non-interacting resonant level model with one lead and compute population of the dot at finite times. This is done in the next section.

## 2. Relaxation of an observable at finite times

Consider non-interacting resonant level model

$$\hat{H} = \sum_k \varepsilon_k \hat{c}_k^\dagger \hat{c}_k + \varepsilon_d \hat{d}^\dagger \hat{d} + \sum_k \left( V_k^* \hat{d}^\dagger \hat{c}_k + V_k \hat{d} \hat{c}_k^\dagger \right) \tag{2.1}$$

The density matrix:

$$\hat{\rho}_0 = \frac{1}{Z} \prod_k \exp\left\{ -\beta_\alpha (\varepsilon_k - \mu) \hat{\psi}_k^\dagger \hat{\psi}_k \right\} \tag{2.2}$$

After the sudden kick (t=0) $\varepsilon_d \to \varepsilon_{d2}$:

$$\hat{H}_2 = \sum_k \varepsilon_k \hat{c}_k^\dagger \hat{c}_k + \varepsilon_{d2} \hat{d}^\dagger \hat{d} + \sum_k \left( V_k^* \hat{d}^\dagger \hat{c}_k + V_k \hat{d} \hat{c}_k^\dagger \right) \tag{2.3}$$

Time evolution of $\langle \hat{d}^\dagger \hat{d} \rangle$ over time:

$$\langle \hat{d}^\dagger \hat{d} \rangle = Tr\{ \hat{d}^\dagger \hat{d} \exp(-iH_2 t) \hat{\rho}_0 \exp(iH_2 t) \} \tag{2.4}$$

First, we need to express $\hat{d}^\dagger \hat{d}$ through the new steady state scattering operators, index 2 corresponds to the green's functions and scattering states of (2.3):

$$\hat{d}^\dagger \hat{d} = \sum_{kn} G_{dd2}^a(\varepsilon_k) V_k G_{dd2}^r(\varepsilon_n) V_n^* \hat{\psi}_{k2}^\dagger \hat{\psi}_{n2} \tag{2.5}$$

Taking into account that

$$(\hat{\psi}_{k2}^\dagger - \hat{c}_n^\dagger) / G_{dd2}^r(\varepsilon_k) = (\hat{\psi}_k^\dagger - \hat{c}_n^\dagger) / G_{dd}^r(\varepsilon_k) \tag{2.6}$$

we have

$$\hat{\psi}_{n2} = \hat{\psi}_n + \{1 - G^a_{dd2}(\varepsilon_n)/G^a_{dd}(\varepsilon_n)\} \sum_m V_n \frac{V_m^* G^r_{dd}(\varepsilon_m)}{\varepsilon_m - \varepsilon_n + i\eta} \hat{\psi}_m$$
$$= \hat{\psi}_n + G^a_{dd2}(\varepsilon_n)(\varepsilon_d - \varepsilon_{d2}) \sum_m V_n \frac{V_m^* G^r_{dd}(\varepsilon_m)}{\varepsilon_m - \varepsilon_n + i\eta} \hat{\psi}_m \qquad (2.7)$$

$$\hat{\psi}^\dagger_{k2} = \hat{\psi}^\dagger_k + \{1 - G^r_{dd2}(\varepsilon_k)/G^r_{dd}(\varepsilon_k)\} \sum_m V_k^* \frac{V_m G^a_{dd}(\varepsilon_m)}{\varepsilon_m - \varepsilon_k - i\eta} \hat{\psi}^\dagger_m$$
$$= \hat{\psi}^\dagger_k + G^r_{dd2}(\varepsilon_k)(\varepsilon_d - \varepsilon_{d2}) \sum_m V_k^* \frac{V_m G^a_{dd}(\varepsilon_m)}{\varepsilon_m - \varepsilon_k - i\eta} \hat{\psi}^\dagger_m \qquad (2.8)$$

Taking into account that

$$\exp(iH_2 t) \hat{\psi}^\dagger_{k2} \hat{\psi}_{n2} \exp(-iH_2 t) = \hat{\psi}^\dagger_{k2} \hat{\psi}_{n2} \exp(i(\varepsilon_k - \varepsilon_n)t) \qquad (2.9)$$

we have for (2.4)

$$\langle \hat{d}^\dagger \hat{d} \rangle = Tr\{\hat{d}^\dagger \hat{d} \exp(-iH_2 t) \hat{\rho}_0 \exp(iH_2 t)\}$$
$$= \sum_{kn} \exp(i(\varepsilon_k - \varepsilon_n)t) G^a_{dd2}(\varepsilon_k) V_k G^r_{dd2}(\varepsilon_n) V_n^* Tr\{\hat{\psi}^\dagger_{k2} \hat{\psi}_{n2} \hat{\rho}_0\} \qquad (2.10)$$

And with (2.7) and (2.8):

$$Tr\{\hat{\psi}^\dagger_{k2} \hat{\psi}_{n2} \hat{\rho}_0\}$$
$$= \delta_{kn} f(\varepsilon_n)$$
$$+ G^r_{dd2}(\varepsilon_k)(\varepsilon_d - \varepsilon_{d2}) V_k^* \frac{V_n G^a_{dd}(\varepsilon_n)}{\varepsilon_n - \varepsilon_k - i\eta} f(\varepsilon_n) + G^a_{dd2}(\varepsilon_n)(\varepsilon_d - \varepsilon_{d2}) V_n \frac{V_k^* G^r_{dd}(\varepsilon_k)}{\varepsilon_k - \varepsilon_n + i\eta} f(\varepsilon_k) \quad (2.11)$$
$$+ G^r_{dd2}(\varepsilon_k) G^a_{dd2}(\varepsilon_n)(\varepsilon_d - \varepsilon_{d2})^2 \sum_m V_n \frac{V_m^* G^r_{dd}(\varepsilon_m)}{\varepsilon_m - \varepsilon_n + i\eta} V_k^* \frac{V_m G^a_{dd}(\varepsilon_m)}{\varepsilon_m - \varepsilon_k - i\eta} f(\varepsilon_m)$$

Inserting (2.11) into (2.10):

$$\langle \hat{d}^\dagger \hat{d} \rangle$$

$$= \sum_{kn} \delta_{kn} f(\varepsilon_n) \exp(i(\varepsilon_k - \varepsilon_n)t) G^a_{dd2}(\varepsilon_k) V_k V_n^* G^r_{dd2}(\varepsilon_n)$$

$$+ \sum_{kn} \exp(i(\varepsilon_k - \varepsilon_n)t) G^a_{dd2}(\varepsilon_k) V_k V_n^* G^r_{dd2}(\varepsilon_n) G^r_{dd2}(\varepsilon_k)(\varepsilon_d - \varepsilon_{d2}) V_k^* \frac{V_n G^a_{dd}(\varepsilon_n)}{\varepsilon_n - \varepsilon_k - i\eta} f(\varepsilon_n)$$

$$+ \sum_{kn} \exp(i(\varepsilon_k - \varepsilon_n)t) G^a_{dd2}(\varepsilon_k) V_k V_n^* G^r_{dd2}(\varepsilon_n) G^a_{dd2}(\varepsilon_n)(\varepsilon_d - \varepsilon_{d2}) V_n \frac{V_k^* G^r_{dd}(\varepsilon_k)}{\varepsilon_k - \varepsilon_n + i\eta} f(\varepsilon_k)$$

$$+ \sum_{kn} \exp(i(\varepsilon_k - \varepsilon_n)t) G^a_{dd2}(\varepsilon_k) V_k G^r_{dd2}(\varepsilon_n) V_n^* G^r_{dd2}(\varepsilon_k) G^a_{dd2}(\varepsilon_n)(\varepsilon_d - \varepsilon_{d2})^2 \sum_m V_n \frac{V_m^* G^r_{dd}(\varepsilon_m)}{\varepsilon_m - \varepsilon_n + i\eta} V_k^* \frac{V_m G^a_{dd}(\varepsilon_m)}{\varepsilon_m - \varepsilon_k - i\eta} f(\varepsilon_m)$$

(2.12)

This expressing has 4 large terms. Let's evaluate them:

$$\sum_{kn} \delta_{kn} f(\varepsilon_n) \exp(i(\varepsilon_k - \varepsilon_n)t) G^a_{dd2}(\varepsilon_k) V_k V_n^* G^r_{dd2}(\varepsilon_n)$$

$$= \sum_k f(\varepsilon_k) G^a_{dd2}(\varepsilon_k) V_k V_k^* G^r_{dd2}(\varepsilon_k) = \frac{1}{2\pi} \int_{-\infty}^{\infty} G^a_{dd2}(\varepsilon) G^r_{dd2}(\varepsilon) \Gamma(\varepsilon) f(\varepsilon) d\varepsilon \qquad (2.13)$$

$$= \frac{1}{2\pi} \int_{-\infty}^{\infty} A_{dd2} f d\varepsilon$$

(reducing the summation to the integral is a procedure used throughout the manuscript and not showed here for the sake of brevity)

The second term:

$$\sum_{kn} \exp(i(\varepsilon_k - \varepsilon_n)t) G^a_{dd2}(\varepsilon_k) V_k V_n^* G^r_{dd2}(\varepsilon_n) G^r_{dd2}(\varepsilon_k)(\varepsilon_d - \varepsilon_{d2}) V_k^* \frac{V_n G^a_{dd}(\varepsilon_n)}{\varepsilon_n - \varepsilon_k - i\eta} f(\varepsilon_n)$$

$$= \left(\frac{1}{2\pi}\right)^2 (\varepsilon_d - \varepsilon_{d2}) \int_{-\infty}^{\infty} f(\varepsilon') \Gamma(\varepsilon') d\varepsilon' G^a_{dd}(\varepsilon') G^r_{dd2}(\varepsilon') \int_{-\infty}^{\infty} \exp(i(\varepsilon - \varepsilon')t) G^a_{dd2}(\varepsilon) G^r_{dd2}(\varepsilon) \Gamma(\varepsilon) \frac{1}{\varepsilon' - \varepsilon - i\eta} d\varepsilon$$

(2.14)

Before proceeding further let's evaluate the following integral:

$$\int_{-\infty}^{\infty} A_{dd}(\varepsilon) \frac{1}{\varepsilon' - \varepsilon - i\eta} d\varepsilon \qquad (2.15)$$

To do that, we convert the integration to the summation:

$$\left(\frac{1}{2\pi}\right)\int_{-\infty}^{\infty} A_{dd}(\varepsilon)\frac{1}{\varepsilon'-\varepsilon-i\eta}d\varepsilon = \sum_{k} G_{dd}^{a}(\varepsilon_k)V_k V_k^* G_{dd}^{r}(\varepsilon_k)\frac{1}{\varepsilon'-\varepsilon_k-i\eta}$$

$$= \langle d|\sum_{k} \frac{|\psi_k\rangle\langle\psi_k|}{\varepsilon'-\varepsilon_k-i\eta}|d\rangle = \langle d|G^a(\varepsilon')|d\rangle = G_{dd}^a(\varepsilon') \quad (2.16)$$

Using (2.16) one can perform "sanity check" for t=0: we have for (2.14) at t=0:

$$\left(\frac{1}{2\pi}\right)(\varepsilon_d - \varepsilon_{d2})\int_{-\infty}^{\infty} f(\varepsilon')\Gamma(\varepsilon')G_{dd}^a(\varepsilon')G_{dd2}^r(\varepsilon')G_{dd2}^a(\varepsilon')d\varepsilon'$$

$$= \left(\frac{1}{2\pi}\right)(\varepsilon_d - \varepsilon_{d2})\int_{-\infty}^{\infty} f(\varepsilon')G_{dd}^a(\varepsilon')A_{dd2}(\varepsilon')d\varepsilon' \quad (2.17)$$

By analogy the third term:

$$\left(\frac{1}{2\pi}\right)(\varepsilon_d - \varepsilon_{d2})\int_{-\infty}^{\infty} f(\varepsilon')G_{dd}^r(\varepsilon')A_{dd2}(\varepsilon')d\varepsilon' \quad (2.18)$$

And the last term:

$$\sum_{kn} G_{dd2}^a(\varepsilon_k)V_k G_{dd2}^r(\varepsilon_n)V_n^* G_{dd2}^r(\varepsilon_k)G_{dd2}^a(\varepsilon_n)(\varepsilon_d - \varepsilon_{d2})^2 \sum_{m} V_n \frac{V_m^* G_{dd}^r(\varepsilon_m)}{\varepsilon_m - \varepsilon_n + i\eta} V_k^* \frac{V_m G_{dd}^a(\varepsilon_m)}{\varepsilon_m - \varepsilon_k - i\eta} f(\varepsilon_m)$$

$$= (\varepsilon_d - \varepsilon_{d2})^2 \sum_{m} V_m^* G_{dd}^r(\varepsilon_m)V_m G_{dd}^a(\varepsilon_m) f(\varepsilon_m) G_{dd2}^a(\varepsilon_m) G_{dd2}^r(\varepsilon_m)$$

$$= \left(\frac{1}{2\pi}\right)(\varepsilon_{d2} - \varepsilon_d)^2 \int_{-\infty}^{\infty} f(\varepsilon')\frac{1}{(\varepsilon'-\varepsilon_d-\Lambda(\varepsilon'))^2 + \Gamma^2(\varepsilon')/4} A_{dd2}(\varepsilon')d\varepsilon'$$

(2.19)

Summation of (2.19),(2.18),(2.17) and (2.13) leads to:

$$\left(\frac{1}{2\pi}\right)\int_{-\infty}^{\infty} f(\varepsilon')A_{dd2}(\varepsilon')\{\frac{(\varepsilon_{d2}-\varepsilon_d)^2}{(\varepsilon'-\varepsilon_d-\Lambda(\varepsilon'))^2 + \Gamma^2(\varepsilon')/4} + \frac{2(\varepsilon_{d2}-\varepsilon_d)(\varepsilon'-\varepsilon_d-\Lambda(\varepsilon'))}{(\varepsilon'-\varepsilon_d-\Lambda(\varepsilon'))^2 + \Gamma^2(\varepsilon')/4} + 1\}d\varepsilon'$$

$$= \left(\frac{1}{2\pi}\right)\int_{-\infty}^{\infty} f(\varepsilon')A_{dd2}(\varepsilon')\{\frac{(\varepsilon_{d2}-\varepsilon_d)^2 + 2(\varepsilon_{d2}-\varepsilon_d)(\varepsilon'-\varepsilon_d-\Lambda(\varepsilon')) + (\varepsilon'-\varepsilon_d)^2 + \Gamma^2(\varepsilon')/4}{(\varepsilon'-\varepsilon_d-\Lambda(\varepsilon'))^2 + \Gamma^2(\varepsilon')/4}\}d\varepsilon'$$

$$= \left(\frac{1}{2\pi}\right)\int_{-\infty}^{\infty} f(\varepsilon')A_{dd2}(\varepsilon')\{\frac{(\varepsilon'-\varepsilon_{d2}-\Lambda(\varepsilon'))^2 + \Gamma^2(\varepsilon')/4}{(\varepsilon'-\varepsilon_d-\Lambda(\varepsilon'))^2 + \Gamma^2(\varepsilon')/4}\}d\varepsilon'$$

$$= \left(\frac{1}{2\pi}\right)\int_{-\infty}^{\infty} f(\varepsilon')A_{dd}(\varepsilon')d\varepsilon'$$

(2.20)

which is the equilibrium population of the dot as expected at t = 0 ( immediately after the kick the relaxation has not started yet)

At times t>0 we assume the wide band approximation in odder tot analytically evaluate time-dependent exponents.

Then let's evaluate the following integral:

$$I(t) = \frac{1}{2\pi} \int_{-\infty}^{\infty} \exp(i(\varepsilon - \varepsilon')t) A_{dd2}(\varepsilon) \frac{1}{\varepsilon' - \varepsilon - i\eta} d\varepsilon \qquad (2.21)$$

We differentiate (2.21) with respect to time:

$$\partial_t I(t) = \frac{1}{2\pi} \int_{-\infty}^{\infty} \exp(i(\varepsilon - \varepsilon')t) A_{dd2} \frac{i(\varepsilon - \varepsilon')}{\varepsilon' - \varepsilon - i\eta} d\varepsilon$$

$$= -\frac{1}{2\pi} \int_{-\infty}^{\infty} i \exp(i(\varepsilon - \varepsilon')t) A_{dd2} d\varepsilon = \qquad (2.22)$$

$$= -\frac{1}{2\pi} \int_{-\infty}^{\infty} i \exp(i(\varepsilon - \varepsilon')t) \frac{\Gamma}{(\varepsilon - \varepsilon_{d2})^2 + \Gamma^2/4} d\varepsilon$$

The residue points:

$$\varepsilon_{\pm} = \varepsilon_d \pm i\Gamma/2 \qquad (2.23)$$

which means that:

$$\partial_t I(t) = -\frac{1}{2\pi} \int_{-\infty}^{\infty} i \exp(i(\varepsilon - \varepsilon')t) \frac{\Gamma}{(\varepsilon - \varepsilon_{d2})^2 + \Gamma^2/4} d\varepsilon$$

$$= -\frac{1}{2\pi} i 2\pi \exp(i(\varepsilon_{d2} - \varepsilon')t - \Gamma/2t) = -i \exp(i(\varepsilon_{d2} - \varepsilon')t - \Gamma/2t) \qquad (2.24)$$

Thus

$$I(t) = C - \frac{\exp(i(\varepsilon_d - \varepsilon')t - \Gamma/2t)}{(\varepsilon_{d2} - \varepsilon') + i\Gamma/2} = C + \frac{\exp(i(\varepsilon_d - \varepsilon')t - \Gamma/2t)}{\varepsilon' - \varepsilon_{d2} - i\Gamma/2} \qquad (2.25)$$

With the boundary condition (2.16) we have:

$$I(t) = G_{dd2}^a(\varepsilon') \exp(i(\varepsilon_{d2} - \varepsilon')t - \Gamma/2t) \qquad (2.26)$$

With (2.26) the second term (2.14) becomes:

$$\left(\frac{1}{2\pi}\right)(\varepsilon_d - \varepsilon_{d2})\int_{-\infty}^{\infty} f(\varepsilon')\Gamma G_{dd}^a(\varepsilon')G_{dd2}^r(\varepsilon')G_{dd2}^a(\varepsilon')\exp(i(\varepsilon_{d2}-\varepsilon')t - \Gamma/2t)d\varepsilon'$$

$$= \left(\frac{1}{2\pi}\right)(\varepsilon_d - \varepsilon_{d2})\exp(-\Gamma/2t)\int_{-\infty}^{\infty} f(\varepsilon')G_{dd}^a(\varepsilon')A_{dd2}(\varepsilon')\exp(i(\varepsilon_{d2}-\varepsilon')t)d\varepsilon' \tag{2.27}$$

The third term becomes:

$$\left(\frac{1}{2\pi}\right)(\varepsilon_d - \varepsilon_{d2})\int_{-\infty}^{\infty} f(\varepsilon')\Gamma G_{dd}^a(\varepsilon')G_{dd2}^r(\varepsilon')G_{dd2}^r(\varepsilon')\exp(-i(\varepsilon_{d2}-\varepsilon')t - \Gamma/2t)d\varepsilon' =$$

$$= \left(\frac{1}{2\pi}\right)(\varepsilon_d - \varepsilon_{d2})\exp(-\Gamma/2t)\int_{-\infty}^{\infty} f(\varepsilon')G_{dd}^r(\varepsilon')A_{dd2}(\varepsilon')\exp(-i(\varepsilon_{d2}-\varepsilon')t)d\varepsilon' \tag{2.28}$$

The fourth term

$$\sum_{kn}\exp(i(\varepsilon_k-\varepsilon_n)t)G_{dd2}^a(\varepsilon_k)V_k G_{dd2}^r(\varepsilon_n)V_n^* G_{dd2}^r(\varepsilon_k)G_{dd2}^a(\varepsilon_n)(\varepsilon_d-\varepsilon_{d2})^2 \sum_m V_n \frac{V_m^* G_{dd}^r(\varepsilon_m)}{\varepsilon_m-\varepsilon_n+i\eta}V_k^* \frac{V_m G_{dd}^a(\varepsilon_m)}{\varepsilon_m-\varepsilon_k-i\eta}f(\varepsilon_m)$$

$$= \sum_{kn} G_{dd2}^a(\varepsilon_k)V_k G_{dd2}^r(\varepsilon_n)V_n^* G_{dd2}^r(\varepsilon_k)G_{dd2}^a(\varepsilon_n)(\varepsilon_d-\varepsilon_{d2})^2$$

$$\sum_m V_n \frac{V_m^* G_{dd}^r(\varepsilon_m)}{\varepsilon_m-\varepsilon_n+i\eta}V_k^* \frac{V_m G_{dd}^a(\varepsilon_m)}{\varepsilon_m-\varepsilon_k-i\eta}f(\varepsilon_m)\exp(i(\varepsilon_k-\varepsilon_m)t)\exp(i(\varepsilon_m-\varepsilon_n)t)$$

(2.29)

In Eq. (2.29) we perform summation over $k$ and $n$ by turning them into integrals:

for k dependent part:

$$\sum_k G_{dd2}^a(\varepsilon_k)V_k G_{dd2}^r(\varepsilon_k)V_k^* \frac{\exp(i(\varepsilon_k-\varepsilon_m)t)}{\varepsilon_m-\varepsilon_k-i\eta}$$

$$= \frac{1}{2\pi}\int_{-\infty}^{\infty} \exp(i(\varepsilon-\varepsilon_m)t)A_{dd2}(\varepsilon)\frac{1}{\varepsilon_m-\varepsilon-i\eta}d\varepsilon \tag{2.30}$$

$$= G_{dd2}^a(\varepsilon_m)\exp(i(\varepsilon_{d2}-\varepsilon_m)t - \Gamma/2t)$$

for n dependent part:

$$\sum_n G_{dd2}^a(\varepsilon_n)V_n G_{dd2}^r(\varepsilon_n)V_n^* \frac{\exp(i(\varepsilon_m-\varepsilon_n)t)}{\varepsilon_m-\varepsilon_n+i\eta}$$

$$= G_{dd2}^r(\varepsilon_m)\exp(-i(\varepsilon_{d2}-\varepsilon_m)t - \Gamma/2t) \tag{2.31}$$

And combining (2.31), (2.30) with (2.29) we have for the fourth term:

$$\left(\frac{1}{2\pi}\right)(\varepsilon_d-\varepsilon_{d2})^2 \exp(-\Gamma t)\int_{-\infty}^{\infty} f(\varepsilon')\frac{1}{(\varepsilon-\varepsilon_d)^2+\Gamma^2/4}A_{dd2}(\varepsilon')d\varepsilon' \tag{2.32}$$

Gathering (2.32),(2.28), (2.27) and (2.13) we finally have :

$$\langle \hat{d}^\dagger \hat{d} \rangle$$

$$= \frac{1}{2\pi} \int_{-\infty}^{\infty} A_{dd2} f d\varepsilon$$

$$+ \left(\frac{1}{2\pi}\right)(\varepsilon_d - \varepsilon_{d2}) \exp(-\Gamma/2t) \int_{-\infty}^{\infty} f A_{dd2} \{G_{dd}^r \exp(-i(\varepsilon_{d2} - \varepsilon)t) + G_{dd}^a \exp(i(\varepsilon_{d2} - \varepsilon)t)\} d\varepsilon$$

$$+ \left(\frac{1}{2\pi}\right)(\varepsilon_d - \varepsilon_{d2})^2 \exp(-\Gamma t) \int_{-\infty}^{\infty} f \frac{1}{(\varepsilon - \varepsilon_d)^2 + \Gamma^2/4} A_{dd2}(\varepsilon) d\varepsilon$$

$$= \frac{1}{2\pi} \int_{-\infty}^{\infty} A_{dd2} f d\varepsilon$$

$$+ \left(\frac{1}{2\pi}\right)(\varepsilon_d - \varepsilon_{d2}) \exp(-\Gamma/2t) \int_{-\infty}^{\infty} f A_{dd2} \frac{2(\varepsilon - \varepsilon_d)\cos((\varepsilon_{d2} - \varepsilon)t) - \Gamma \sin((\varepsilon_{d2} - \varepsilon)t)}{(\varepsilon - \varepsilon_d)^2 + \Gamma^2/4} d\varepsilon$$

$$+ \left(\frac{1}{2\pi}\right)(\varepsilon_d - \varepsilon_{d2})^2 \exp(-\Gamma t) \int_{-\infty}^{\infty} f \frac{1}{(\varepsilon - \varepsilon_d)^2 + \Gamma^2/4} A_{dd2} d\varepsilon \tag{2.33}$$

As expected that $t \to \infty$ $\langle \hat{d}^\dagger \hat{d} \rangle$ becomes $\frac{1}{2\pi} \int_{-\infty}^{\infty} A_{dd2} f d\varepsilon$ which is an equilibrium dot population at new position $\varepsilon_{d2}$ of the dot level. It is also easy to check that at t=0:

$$\langle \hat{d}^\dagger \hat{d} \rangle$$

$$= \frac{1}{2\pi} \int_{-\infty}^{\infty} A_{dd2} f d\varepsilon$$

$$+ \left(\frac{1}{2\pi}\right)(\varepsilon_d - \varepsilon_{d2}) \int_{-\infty}^{\infty} f A_{dd2} \frac{2(\varepsilon - \varepsilon_d)}{(\varepsilon - \varepsilon_d)^2 + \Gamma^2/4} d\varepsilon$$

$$+ \left(\frac{1}{2\pi}\right)(\varepsilon_d - \varepsilon_{d2})^2 \int_{-\infty}^{\infty} f \frac{1}{(\varepsilon - \varepsilon_d)^2 + \Gamma^2/4} A_{dd2} d\varepsilon$$

$$= \frac{1}{2\pi} \int_{-\infty}^{\infty} A_{dd2} f (1 + \frac{2(\varepsilon - \varepsilon_d)(\varepsilon_d - \varepsilon_{d2})}{(\varepsilon - \varepsilon_d)^2 + \Gamma^2/4} + \frac{(\varepsilon_d - \varepsilon_{d2})^2}{(\varepsilon - \varepsilon_d)^2 + \Gamma^2/4}) d\varepsilon$$

$$= \frac{1}{2\pi} \int_{-\infty}^{\infty} A_{dd2} f (\frac{2(\varepsilon - \varepsilon_d)(\varepsilon_d - \varepsilon_{d2}) + (\varepsilon_d - \varepsilon_{d2})^2 + (\varepsilon - \varepsilon_d)^2 + \Gamma^2/4}{(\varepsilon - \varepsilon_d)^2 + \Gamma^2/4}) d\varepsilon$$

$$= \frac{1}{2\pi} \int_{-\infty}^{\infty} A_{dd2} f (\frac{(\varepsilon - \varepsilon_{d2})^2 + \Gamma^2/4 + \Gamma^2/4}{(\varepsilon - \varepsilon_d)^2 + \Gamma^2/4}) d\varepsilon = \frac{1}{2\pi} \int_{-\infty}^{\infty} A_{dd} f d\varepsilon$$

(2.34)

the dot population is the initial dot population as expected.

By analogy, one can compute relaxation of any observables.

We can also check that (2.33) is the same expression obtained in Ref 2.

Indeed, in this paper:

$$\langle \hat{d}^\dagger \hat{d} \rangle = \frac{\Gamma}{2\pi} \int_{-\infty}^{\infty} |A|^2 f d\varepsilon \qquad (2.35)$$

where

$$A = \frac{1}{\varepsilon - \varepsilon_d + i\Gamma/2} \{1 + \frac{(\varepsilon_{d2} - \varepsilon_d)(1 - \exp\{i(\varepsilon - \varepsilon_{d2}) - \Gamma/2)t\})}{\varepsilon - \varepsilon_{d2} + i\Gamma/2}\} \qquad (2.36)$$

We re-write (2.36) as follows:

$$A = \frac{1}{\varepsilon - \varepsilon_d + i\Gamma/2} \{1 + \frac{(\varepsilon_{d2} - \varepsilon_d)(1 - \exp\{i(\varepsilon - \varepsilon_{d2}) - \Gamma/2)t\})}{\varepsilon - \varepsilon_{d2} + i\Gamma/2}\}$$
$$= \frac{1}{\varepsilon - \varepsilon_d + i\Gamma/2} \{1 + \frac{(\varepsilon_{d2} - \varepsilon_d)}{\varepsilon - \varepsilon_{d2} + i\Gamma/2} - \frac{(\varepsilon_{d2} - \varepsilon_d)\exp\{i(\varepsilon - \varepsilon_{d2}) - \Gamma/2)t\}}{\varepsilon - \varepsilon_{d2} + i\Gamma/2}\} \qquad (2.37)$$
$$= \frac{1}{\varepsilon - \varepsilon_{d2} + i\Gamma/2}(1 - \frac{(\varepsilon_{d2} - \varepsilon_d)\exp\{i(\varepsilon - \varepsilon_{d2}) - \Gamma/2)t\}}{\varepsilon - \varepsilon_d + i\Gamma/2})$$

Thus

$$\Gamma|A|^2 = A_{dd2}(1 - (\varepsilon_{d2} - \varepsilon_d)\frac{\exp\{i(\varepsilon - \varepsilon_{d2}) - \Gamma/2)t\}}{\varepsilon - \varepsilon_d + i\Gamma/2})(1 - (\varepsilon_{d2} - \varepsilon_d)\frac{\exp\{-i(\varepsilon - \varepsilon_{d2}) - \Gamma/2)t\}}{\varepsilon - \varepsilon_d - i\Gamma/2})$$
$$= A_{dd2}(1 + (\varepsilon_d - \varepsilon_{d2})\frac{2(\varepsilon - \varepsilon_d)\cos((\varepsilon_{d2} - \varepsilon)t) - \Gamma\sin((\varepsilon_{d2} - \varepsilon)t)}{(\varepsilon - \varepsilon_d)^2 + \Gamma^2/4}\exp(-\Gamma/2 t)$$
$$+ (\varepsilon_d - \varepsilon_{d2})^2 \frac{1}{(\varepsilon - \varepsilon_d)^2 + \Gamma^2/4}\exp(-\Gamma t))$$

(2.38)

which, after substituting in (2.35), coincides with (2.33).

### 3. Currents

Eqs. (2.7) - (2.8) are written for the incoming solution ( obtained using the Lipmann-Schwinger Equation for the retarded Green's function, see the manuscript). For the outgoing solution we use the advanced GF, which gives

$$\hat{\psi}_{n2,-} = \hat{\psi}_{n,-} + G^r_{dd2}(\varepsilon_n)(\varepsilon_d - \varepsilon_{d2})\sum_m V_n \frac{V_m^* G^a_{dd}(\varepsilon_m)}{\varepsilon_m - \varepsilon_n - i\eta}\hat{\psi}_{m,-} \qquad (3.1)$$

$$\hat{\psi}_{k2,-} = \hat{\psi}_{k,-} + G_{dd2}^a(\varepsilon_k)(\varepsilon_d - \varepsilon_{d2})\sum_m V_k^* \frac{V_m G_{dd}^r(\varepsilon_m)}{\varepsilon_m - \varepsilon_k + i\eta}\hat{\psi}_{m,-} \quad (3.2)$$

Eqs. (3.1) and (3.2) refer to the creation/annihilation of a scattering wave function (1D plane, since the problem is 1D). To connect our formalism to the Landauer-Buttiker description, recall that the total field operator:

$$\hat{\chi}_{k,\pm}^\dagger(t) = \frac{1}{\sqrt{2\pi v_k}}\hat{\psi}_{k,\pm}^\dagger \sqrt{D_k} \quad (3.3)$$

where $p_k$ and $v_k$ are momentum and velocity, $\hat{\psi}_{k,+}^\dagger$ is just $\hat{\psi}_k^\dagger$ (retarded solution) used above in section 2 and $D_k = \sum_n 2\pi\delta(p_k - p_n)$ is the density of states in the momentum space. One can see that

$$Tr\{\hat{\rho}_0 \hat{\chi}_{k,+}^\dagger \hat{\chi}_{n,+}\} = \frac{1}{\sqrt{2\pi v_n}}\frac{1}{\sqrt{2\pi v_k}} f(\varepsilon_k)\delta_{kn} 2\pi\delta(p_k - p_n)$$

$$= \frac{1}{v_k}\partial_{p_k}\varepsilon_k f(\varepsilon_k)\delta_{kn} 2\pi\delta(\varepsilon_k - \varepsilon_k) = \frac{p_k}{mv_k}f(\varepsilon_k)\delta_{kn}\delta(\varepsilon_k - \varepsilon_n) = f(\varepsilon_k)\delta_{kn}\delta(\varepsilon_k - \varepsilon_n) \quad (3.4)$$

The total incoming flux of particles:

$$I_{inc} = \frac{1}{2\pi}\int_{-\infty}^{\infty} d\varepsilon_k \int_{-\infty}^{\infty} d\varepsilon_n Tr\{\hat{\rho}_0 \hat{\chi}_{k,+}^\dagger \hat{\chi}_{n,+}\} \quad (3.5)$$

and the outgoing flux

$$I_{out} = \frac{1}{2\pi}\int_{-\infty}^{\infty} d\varepsilon_k \int_{-\infty}^{\infty} d\varepsilon_n Tr\{\hat{\rho}_0 \hat{\chi}_{k,-}^\dagger \hat{\chi}_{n,-}\} \quad (3.6)$$

In the equilibrium

$Tr\{\hat{\rho}_0 \hat{\psi}_{k,\pm}^\dagger \hat{\psi}_{n,\pm}\} = f(\varepsilon_k)\delta_{kn}$ (see eq. (J9) in the manuscript), thus

$$Tr\{\hat{\rho}_0 \hat{\chi}_{k,\pm}^\dagger \hat{\chi}_{n,\pm}\} = f(\varepsilon_k)\delta_{kn}\delta(\varepsilon_k - \varepsilon_n) \quad (3.7)$$

and both fluxes (3.6) and (3.5) are equal, thus $\frac{d\langle \hat{d}^\dagger \hat{d}\rangle}{dt} = -I_N = I_{inc} - I_{out} = 0$

In case of the sudden kick outgoing flux:

$$I_{out}(t) = \frac{1}{2\pi}\int_{-\infty}^{\infty} d\varepsilon_k \int_{-\infty}^{\infty} d\varepsilon_n Tr\{\hat{\rho}_0 \hat{\chi}_{k2,-}^\dagger \hat{\chi}_{n2,-}\} \quad (3.8)$$

Consider $Tr\{\hat{\rho}_0 \hat{\chi}^\dagger_{k2,-} \hat{\chi}_{n2,-}\}$:

$Tr\{\hat{\rho}_0 \hat{\chi}^\dagger_{k2,-} \hat{\chi}_{n2,-}\}$

$= f(\varepsilon_k) 2\pi \delta(\varepsilon_k - \varepsilon_n) + G^r_{dd2}(\varepsilon_n)(\varepsilon_d - \varepsilon_{d2}) \sum_m V_n \frac{V_m^* G^a_{dd}(\varepsilon_m)}{\varepsilon_m - \varepsilon_n - i\eta} f(\varepsilon_k) \delta(\varepsilon_k - \varepsilon_m) \exp(i(\varepsilon_n - \varepsilon_k)t)$

$+ G^a_{dd2}(\varepsilon_k)(\varepsilon_d - \varepsilon_{d2}) \sum_m V_k^* \frac{V_m G^r_{dd}(\varepsilon_m)}{\varepsilon_m - \varepsilon_k + i\eta} f(\varepsilon_n) \delta(\varepsilon_n - \varepsilon_m) \exp(i(\varepsilon_n - \varepsilon_k)t)$

$+ G^a_{dd2}(\varepsilon_k)(\varepsilon_d - \varepsilon_{d2}) G^r_{dd2}(\varepsilon_n)(\varepsilon_d - \varepsilon_{d2}) V_k^* V_n \sum_l \frac{V_l G^r_{dd}(\varepsilon_l)}{\varepsilon_l - \varepsilon_k + i\eta} \sum_m \frac{V_m^* G^a_{dd}(\varepsilon_m)}{\varepsilon_m - \varepsilon_n - i\eta} f(\varepsilon_l) \delta(\varepsilon_l - \varepsilon_m) \exp(i(\varepsilon_n - \varepsilon_k)t)$

(3.9)

Since the first term in this expression will be canceled out by the incoming flux, consider the integration of the second term:

$\int_{-\infty}^{\infty} d\varepsilon_k \int_{-\infty}^{\infty} d\varepsilon_n G^r_{dd2}(\varepsilon_n)(\varepsilon_d - \varepsilon_{d2}) \sum_m V_n \frac{V_m^* G^a_{dd}(\varepsilon_m)}{\varepsilon_m - \varepsilon_n - i\eta} f(\varepsilon_k) \delta(\varepsilon_k - \varepsilon_m) \exp(i(\varepsilon_n - \varepsilon_k)t)$

$= \int_{-\infty}^{\infty} d\varepsilon_n G^r_{dd2}(\varepsilon_n)(\varepsilon_d - \varepsilon_{d2}) \sum_m V_n \frac{V_m^* G^a_{dd}(\varepsilon_m)}{\varepsilon_m - \varepsilon_n - i\eta} f(\varepsilon_m) \exp(i(\varepsilon_n - \varepsilon_m)t)$

(3.10)

To evaluate this integral, recall the procedure we used in the section 2: first, we put $t = 0$ to define the boundary conditions and we also assume the wide band approximation in order to evaluate integrals analytically and compare to (2.33):

$\int_{-\infty}^{\infty} d\varepsilon_n G^r_{dd2}(\varepsilon_n)(\varepsilon_d - \varepsilon_{d2}) \sum_m V \frac{V^* G^a_{dd}(\varepsilon_m)}{\varepsilon_m - \varepsilon_n - i\eta} f(\varepsilon_m)$

$= \frac{1}{2\pi} \int_{-\infty}^{\infty} d\varepsilon_n \int_{-\infty}^{\infty} d\varepsilon_m G^r_{dd2}(\varepsilon_n)(\varepsilon_d - \varepsilon_{d2}) \frac{\Gamma G^a_{dd}(\varepsilon_m)}{\varepsilon_m - \varepsilon_n - i\eta} f(\varepsilon_m)$

(3.11)

$= \frac{1}{2\pi}(\varepsilon_d - \varepsilon_{d2}) \int_{-\infty}^{\infty} d\varepsilon_m \Gamma G^a_{dd}(\varepsilon_m) f(\varepsilon_m) \int_{-\infty}^{\infty} d\varepsilon_n G^r_{dd2}(\varepsilon_n) \frac{1}{\varepsilon_m - \varepsilon_n - i\eta}$

To evaluate the expression above recall that

$$\int_{-\infty}^{\infty} d\varepsilon_n G_{dd2}^r(\varepsilon_n) \frac{1}{\varepsilon_m - \varepsilon_n - i\eta}$$

$$= \int_{-\infty}^{\infty} d\varepsilon_n \frac{1}{\varepsilon_n - \varepsilon_{d2} + i\Gamma/2} \frac{1}{\varepsilon_m - \varepsilon_n - i\eta}$$

$$= -\int_{-\infty}^{\infty} d\varepsilon_n \frac{1}{\varepsilon_n - \varepsilon_{d2} + i\Gamma/2} \frac{1}{\varepsilon_n - \varepsilon_m + i\eta} \quad (3.12)$$

$$= \pi i \frac{1}{\varepsilon_m - \varepsilon_{d2} + i\Gamma/2} - \int_{-\infty}^{\infty} d\varepsilon_n \frac{1}{\varepsilon_n - \varepsilon_{d2} + \varepsilon_m + i\Gamma/2} \frac{1}{-\varepsilon_n - \varepsilon_{d2} + \varepsilon_m + i\Gamma/2}$$

$$= 2\pi i \frac{1}{\varepsilon_m - \varepsilon_{d2} + i\Gamma/2}$$

(we used here $PP\int_{-\infty}^{\infty} \frac{1}{x} f(x)dx = \int_0^{\infty} \frac{1}{x}\{f(x) - f(-x)\}dx$ )

Thus for (3.11)

$$i\int_{-\infty}^{\infty} d\varepsilon_m \Gamma G_{dd}^a(\varepsilon_m) f(\varepsilon_m) \frac{1}{\varepsilon_m - \varepsilon_{d2} + i\Gamma/2} = i\int_{-\infty}^{\infty} d\varepsilon_m \Gamma G_{dd}^a(\varepsilon_m) G_{dd2}^r(\varepsilon_m) f(\varepsilon_m) \quad (3.13)$$

For $t > 0$ we differentiate (3.10) with respect to time:

$$\int_{-\infty}^{\infty} d\varepsilon_n G_{dd2}^r(\varepsilon_n)(\varepsilon_d - \varepsilon_{d2}) \sum_m V_n \frac{V_m^* G_{dd}^a(\varepsilon_m)}{\varepsilon_m - \varepsilon_n - i\eta} f(\varepsilon_m) \delta(\varepsilon_k - \varepsilon_m) i(\varepsilon_n - \varepsilon_m) \exp(i(\varepsilon_n - \varepsilon_m)t)$$

$$= \frac{1}{2\pi} \int_{-\infty}^{\infty} d\varepsilon_m \Gamma G_{dd}^a(\varepsilon_m) f(\varepsilon_m) \int_{-\infty}^{\infty} d\varepsilon_n G_{dd2}^r(\varepsilon_n) \frac{i(\varepsilon_n - \varepsilon_m)\exp(i(\varepsilon_n - \varepsilon_m)t)}{\varepsilon_m - \varepsilon_n - i\eta} \quad (3.14)$$

$$= \frac{1}{2\pi} \int_{-\infty}^{\infty} d\varepsilon_m \Gamma G_{dd}^a(\varepsilon_m) f(\varepsilon_m) \int_{-\infty}^{\infty} d\varepsilon_n G_{dd2}^r(\varepsilon_n) i \exp(i(\varepsilon_n - \varepsilon_m)t)$$

$$= \int_{-\infty}^{\infty} d\varepsilon_m \Gamma G_{dd}^a(\varepsilon_m) f(\varepsilon_m) \exp(i(\varepsilon_{d2} - \varepsilon_m)t - \Gamma/2 \cdot t)$$

Comparing (3.13) and (3.14) finally gives us the expression for the second term in (3.9):

$$(\varepsilon_d - \varepsilon_{d2}) i \int_{-\infty}^{\infty} d\varepsilon_m \Gamma G_{dd}^a(\varepsilon_m) G_{dd2}^r(\varepsilon_m) f(\varepsilon_m) \exp(i(\varepsilon_{d2} - \varepsilon_m)t - \Gamma/2 \cdot t) \quad (3.15)$$

(note: we could directly evaluate $\int_{-\infty}^{\infty} d\varepsilon_n G_{dd2}^r(\varepsilon_n) \frac{\exp(i(\varepsilon_n - \varepsilon_m)t)}{\varepsilon_m - \varepsilon_n - i\eta}$ applying several times the residue theorem, however this way is more tedious).

By analogy, the third term:

$$-(\varepsilon_d - \varepsilon_{d2})i \int_{-\infty}^{\infty} d\varepsilon_m \Gamma G_{dd}^r(\varepsilon_m) G_{dd2}^a(\varepsilon_m) f(\varepsilon_m) \exp(-i(\varepsilon_{d2} - \varepsilon_m)t - \Gamma/2t) \quad (3.16)$$

and for the fourth term we twice apply the procedure outlined in Eqs. (3.11)-(3.16) (it is possible because this term can be factorized into two integrals over $\varepsilon_k$ and $\varepsilon_n$, for more details Ref 1)

$$\Gamma(\varepsilon_d - \varepsilon_{d2})^2 \exp(-\Gamma t) \int_{-\infty}^{\infty} f \frac{1}{(\varepsilon_m - \varepsilon_d)^2 + \Gamma^2/4} A_{dd2} d\varepsilon_m \quad (3.17)$$

Thus, combining (3.15), (3.16) and (3.17) for the particle current we have (we have replaced $\varepsilon$ with $\varepsilon_m$):

$$-I_N(t) = I_{inc} - I_{out} = \frac{1}{2\pi}(-\Gamma(\varepsilon_d - \varepsilon_{d2})^2 \exp(-\Gamma t) \int_{-\infty}^{\infty} f \frac{1}{(\varepsilon - \varepsilon_d)^2 + \Gamma^2/4} A_{dd2} d\varepsilon$$

$$+(\varepsilon_d - \varepsilon_{d2})i \int_{-\infty}^{\infty} d\varepsilon \Gamma G_{dd}^r(\varepsilon) G_{dd2}^a(\varepsilon) f(\varepsilon) \exp(-i(\varepsilon_{d2} - \varepsilon)t - \Gamma/2t) \quad (3.18)$$

$$-(\varepsilon_d - \varepsilon_{d2})i \int_{-\infty}^{\infty} d\varepsilon \Gamma G_{dd}^a(\varepsilon) G_{dd2}^r(\varepsilon) f(\varepsilon) \exp(i(\varepsilon_{d2} - \varepsilon)t - \Gamma/2t)\}$$

On the other hand if we differentiate the expression (2.33) with respect to the time variable, we have:

$$\frac{d\langle \hat{d}^\dagger \hat{d} \rangle}{dt} =$$

$$(\varepsilon_d - \varepsilon_{d2}) \exp(-\Gamma/2t) \int_{-\infty}^{\infty} f A_{dd2}(-i(\varepsilon_{d2} - \varepsilon) - \Gamma/2) G_{dd}^r \exp(-i(\varepsilon_{d2} - \varepsilon)t) d\varepsilon$$

$$+\left(\frac{1}{2\pi}\right)(\varepsilon_d - \varepsilon_{d2}) \exp(-\Gamma/2t) \int_{-\infty}^{\infty} f A_{dd2}(i(\varepsilon_{d2} - \varepsilon) - \Gamma/2) G_{dd}^a \exp(i(\varepsilon_{d2} - \varepsilon)t) d\varepsilon$$

$$-\Gamma\left(\frac{1}{2\pi}\right)(\varepsilon_d - \varepsilon_{d2})^2 \exp(-\Gamma t) \int_{-\infty}^{\infty} f \frac{1}{(\varepsilon - \varepsilon_d)^2 + \Gamma^2/4} A_{dd2}(\varepsilon) d\varepsilon$$

$$=\left(\frac{1}{2\pi}\right)\{i \exp(-\Gamma/2t) \int_{-\infty}^{\infty} f G_{dd2}^a G_{dd}^r \exp(-i(\varepsilon_{d2} - \varepsilon)t) d\varepsilon$$

$$-i \exp(-\Gamma/2t) \int_{-\infty}^{\infty} f G_{dd}^a G_{dd2}^r \exp(-(\varepsilon_{d2} - \varepsilon)t) d\varepsilon$$

$$-\Gamma(\varepsilon_d - \varepsilon_{d2})^2 \exp(-\Gamma t) \int_{-\infty}^{\infty} f \frac{1}{(\varepsilon - \varepsilon_d)^2 + \Gamma^2/4} A_{dd2}(\varepsilon) d\varepsilon\} \quad (3.19)$$

which coincides with (3.18) as we expected.

We see that at $t \to \infty$ (3.18) ($I_N(t)$) goes to zero as also expected. We can also re-write (3.18) in a real form:

$$I_N(t) = \frac{1}{2\pi}(\Gamma(\varepsilon_d - \varepsilon_{d2})^2 \exp(-\Gamma t) \int_{-\infty}^{\infty} f \frac{1}{(\varepsilon - \varepsilon_d)^2 + \Gamma^2/4} A_{dd2} d\varepsilon$$
$$+(\varepsilon_{d2} - \varepsilon_d)\exp(-\Gamma t/2) \quad (3.20)$$
$$\times \int_{-\infty}^{\infty} d\varepsilon \{A_{dd2} A_{dd}(\varepsilon_{d2} - \varepsilon_d)f(\varepsilon)\cos((\varepsilon_{d2} - \varepsilon)t) + 2\sin((\varepsilon_{d2} - \varepsilon)t)\Gamma \operatorname{Re}(G^r_{dd}(\varepsilon)G^a_{dd2}(\varepsilon))\}$$

We can also define expressions for the energy currents:

$$IE_{out} = \frac{1}{2\pi} \int_{-\infty}^{\infty} d\varepsilon_k \int_{-\infty}^{\infty} d\varepsilon_n Tr\{\hat{\rho}_0 \hat{\chi}^\dagger_{k,-}(t)\hat{\chi}_{n,-}(t)\}\frac{(\varepsilon_k + \varepsilon_n)}{2} \quad (3.21)$$

$$IE_{in} = \frac{1}{2\pi} \int_{-\infty}^{\infty} d\varepsilon_k \int_{-\infty}^{\infty} d\varepsilon_n Tr\{\hat{\rho}_0 \hat{\chi}^\dagger_{k,+}(t)\hat{\chi}_{n,+}(t)\}\frac{(\varepsilon_k + \varepsilon_n)}{2} \quad (3.22)$$

We can also introduce so called incoming and outgoing distribution (see Ref 3):

$$\varphi_{inc}(\varepsilon,t) = \int_{-\infty}^{\infty} d\Delta Tr\{\hat{\rho}_0 \hat{\chi}^\dagger_{\varepsilon-\Delta/2,+}(t)\hat{\chi}_{\varepsilon+\Delta/2,+}(t)\} \quad (3.23)$$

$$\varphi_{out}(\varepsilon,t) = \int_{-\infty}^{\infty} d\Delta Tr\{\hat{\rho}_0 \hat{\chi}^\dagger_{\varepsilon-\Delta/2,-}(t)\hat{\chi}_{\varepsilon+\Delta/2,-}(t)\} \quad (3.24)$$

And re-write (3.21) and (3.22) as follows ( The Jacobian from $\varepsilon_k$ and $\varepsilon_n$ to $\varepsilon = \frac{(\varepsilon_k + \varepsilon_n)}{2}$ and $\Delta = \varepsilon_n - \varepsilon_k$ is 1)

$$IE_{in} = \frac{1}{2\pi} \int_{-\infty}^{\infty} d\varepsilon \varphi_{in}(\varepsilon,t)\varepsilon \quad (3.25)$$

$$IE_{out} = \frac{1}{2\pi} \int_{-\infty}^{\infty} d\varepsilon \varphi_{out}(\varepsilon,t)\varepsilon \quad (3.26)$$

and for particle currents:

$$I_{out} = \frac{1}{2\pi} \int_{-\infty}^{\infty} d\varepsilon_k \int_{-\infty}^{\infty} d\varepsilon_n Tr\{\hat{\rho}_0 \hat{\chi}^\dagger_{k,-}(t)\hat{\chi}_{n,-}(t)\} = \frac{1}{2\pi} \int_{-\infty}^{\infty} d\varepsilon \varphi_{out}(\varepsilon,t) \quad (3.27)$$

$$I_{in} = \frac{1}{2\pi} \int_{-\infty}^{\infty} d\varepsilon \varphi_{in}(\varepsilon,t) \quad (3.28)$$

One can also define entropy currents:

$$IS_{in/out} = \frac{1}{2\pi} \int_{-\infty}^{\infty} d\varepsilon \{\varphi_{in/out}(\varepsilon,t) \ln \varphi_{in/out}(\varepsilon,t) + (1-\varphi_{in/out}(\varepsilon,t)) \ln(1-\varphi_{in/out}(\varepsilon,t))\} \quad (3.29)$$

and the heat current

$$IQ_{out/in} = \frac{1}{2\pi} \int_{-\infty}^{\infty} (\varepsilon - \mu)\varphi_{out/in}(\varepsilon,t) d\varepsilon = \frac{1}{2\pi} \int_{-\infty}^{\infty} d\varepsilon_k \int_{-\infty}^{\infty} d\varepsilon_n Tr\{\hat{\rho}_0 \hat{\chi}^{\dagger}_{k,\mp}(t) \hat{\chi}_{n,\mp}(t)\} \{\frac{(\varepsilon_k + \varepsilon_n)}{2} - \mu\} \quad (3.30)$$